\documentclass[10pt,journal]{IEEEtranTCOM}

\usepackage[english]{babel}
\usepackage[usenames]{color}
\usepackage[cp1250]{inputenc}
\usepackage{amsfonts}
\usepackage{amsthm}
\usepackage{graphicx}
\usepackage{epsfig}
\usepackage{mathrsfs}
\usepackage{amsmath}
\usepackage{algorithm}
\usepackage{algorithmic}
\usepackage{hyperref}
\DeclareMathOperator*{\argmax}{arg\,max}

\theoremstyle{plain}

\begin{document}
\newcommand{\bea}{\begin{eqnarray}}
\newcommand{\eea}{\end{eqnarray}}
\newcommand{\be}{\begin{equation}}
\newcommand{\ee}{\end{equation}}
\newcommand{\beas}{\begin{eqnarray*}}
\newcommand{\eeas}{\end{eqnarray*}}
\newcommand{\bs}{\backslash}
\newcommand{\bc}{\begin{center}}
\newcommand{\ec}{\end{center}}
\def\SC {\mathscr{C}}

\title{Adaptive stable distribution and Hurst exponent\\ by method of moments moving estimator \\ for nonstationary time series}
\author{\IEEEauthorblockN{Jarek Duda}\\
\IEEEauthorblockA{Jagiellonian University, Institute of Computer science and Computational Mathematics,\\ \L{}ojasiewicza 6,
 Krakow, Poland,
Email: \emph{dudajar@gmail.com}}}
\maketitle

\begin{abstract}
Nonstationarity of real-life time series requires model adaptation. In classical approaches like ARMA-ARCH there is assumed some arbitrarily chosen dependence type. To avoid their bias, we will focus on novel more agnostic approach: moving estimator, which estimates parameters separately for every time $t$: optimizing $F_t=\sum_{\tau<t} (1-\eta)^{t-\tau} \ln(\rho_\theta (x_\tau))$ local log-likelihood with exponentially weakening weights of the old values. In practice such moving estimates can be found by EMA (exponential moving average) of some parameters, like $m_p=E[|x-\mu|^p]$ absolute central moments, updated by $m_{p,t+1} = m_{p,t} + \eta (|x_t-\mu_t|^p-m_{p,t})$. We will focus here on its applications for alpha-Stable distribution, which also influences Hurst exponent, hence can be used for its adaptive estimation. Its application will be shown on financial data as DJIA time series - beside standard estimation of evolution of center $\mu$ and scale parameter $\sigma$, there is also estimated evolution of $\alpha$ parameter allowing to continuously evaluate market stability - tails having $\rho(x) \sim 1/|x|^{\alpha+1}$ behavior, controlling probability of potentially dangerous extreme events.
\end{abstract}
\section{Introduction}
\begin{figure}[t!]
    \centering
        \includegraphics[width=9.cm]{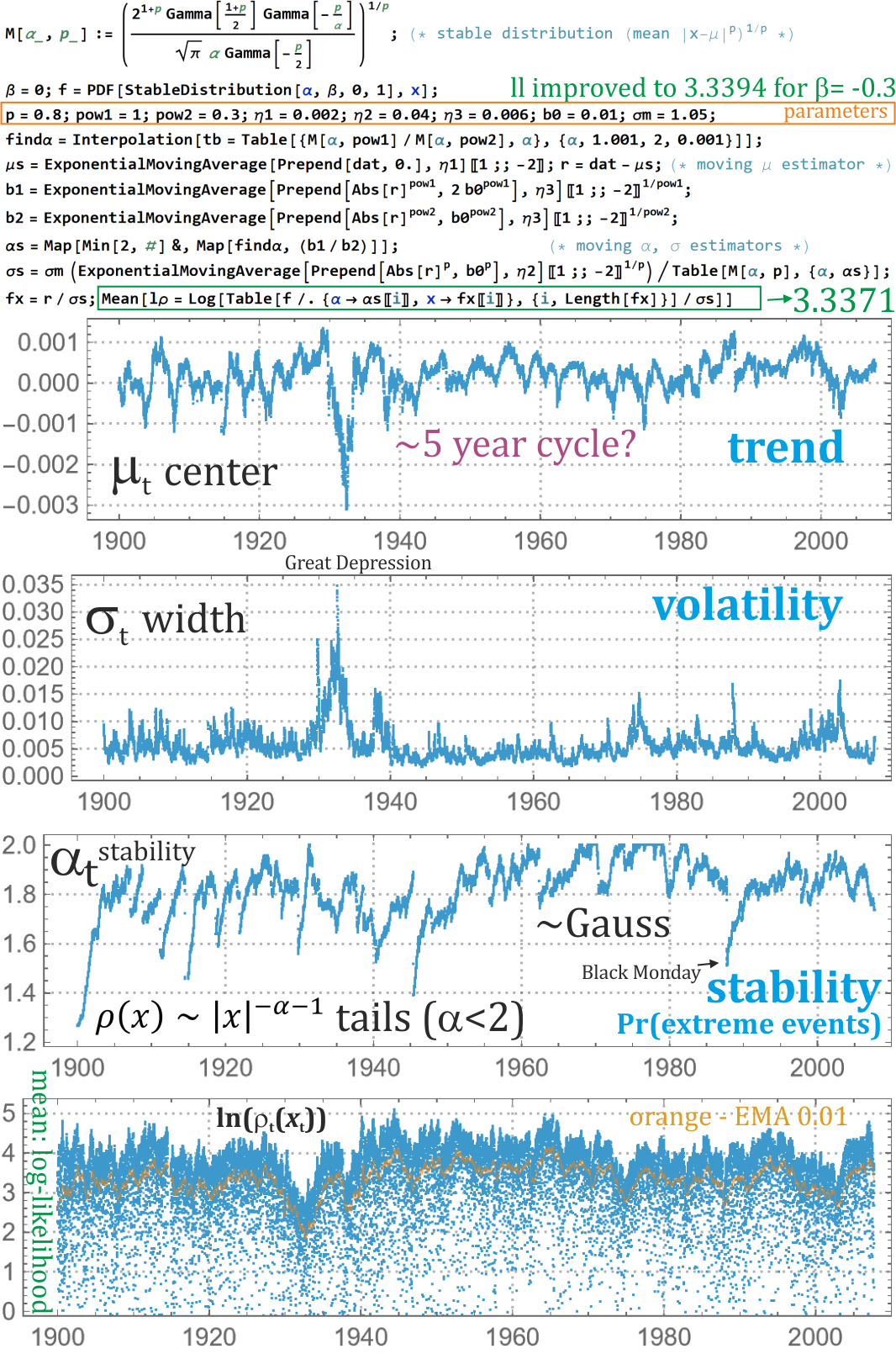}
        \caption{Mathematica code used for moving estimation of  $\theta_t=(\mu_t,\sigma_t,\alpha_t)$ parameters of alpha-stable distribution (for fixed $\beta=0$), using $M_{\nu p}=E[|(x-\mu)/\sigma|^p]$ moment formula (\ref{mom}), and results of its application to 107 years of daily log-returns of DJIA (Dow Jones Industrial Average) time series. The parameters were manually tuned for this case to maximize log-likelihood: mean $\ln(\rho_t(x_t))$ shown at the bottom. Asymmetry by just switching to fixed $\beta=-0.3$ has allowed for its slight improvement. While $\mu$ and $\sigma$ are widely used to evaluate market behavior as \emph{trend} and \emph{volatility}, we can see also interesting evolution of $\alpha$ describing $\rho(x)\sim |x|^{-\alpha-1}$ tail behavior, this way probability of extreme events, we can interpret it as market \emph{stability}, improved especially 1967-1983. }
       \label{djiastab}
\end{figure}

Working with parametric probability distributions, e.g. alpha-stable~(\cite{stable,stablebook}) here, there is a difficult question of choosing its parameters - often done with \textbf{static estimation}: optimization of a single set of parameters, usually through maximization of some function: $F=\frac{1}{T}\sum_{t=1}^T f(\theta,x_t)$, especially log-likelihood in popular MLE (maximal likelihood estimation) using $f(\theta,x)=\ln(\rho_{\theta}(x))$, where $\rho_{\theta}(x)$ is PDF (probability distribution function) for the assumed parametric family. It uses all datapoints with the same $1/T$ contributions, what is appropriate for stationary time series.

\begin{figure*}[t!]
    \centering
        \includegraphics[width=18cm]{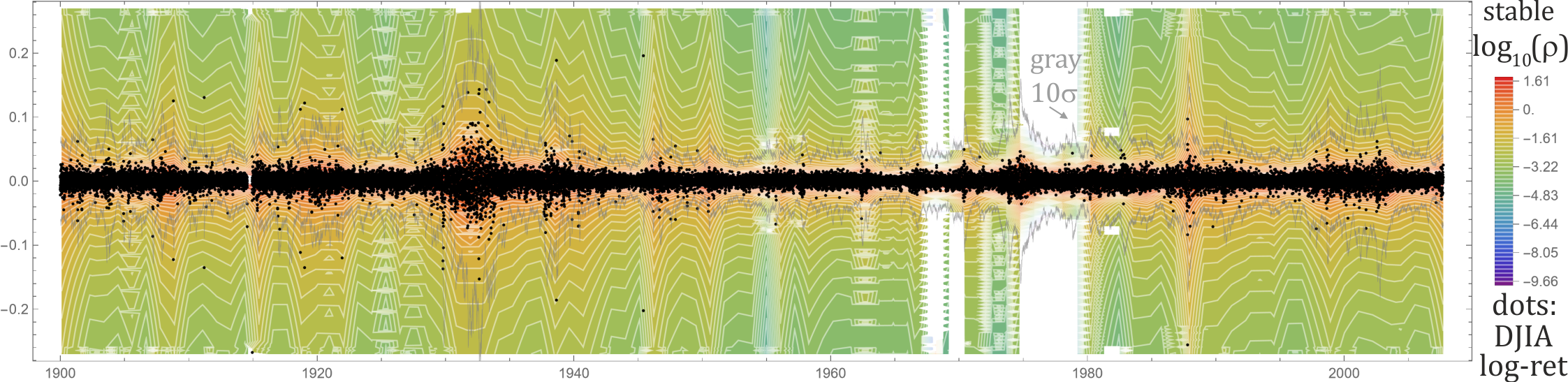}
        \caption{The estimated evolution of probability density for DJIA time series assuming stable distribution, using the code in Fig. \ref{djiastab} with fixed $\beta=-0.3$ asymmetry. The gray plot is estimated $10\sigma$, which for normal distribution is expected to be exceeded once per $\sim 10^{20}$ days, while for DJIA it was exceeded 40 times in $\approx 30000$ days. We can see adaptation makes the tails heavier especially based on observed extreme values.     
         }
       \label{finstab}
\end{figure*}

\begin{figure*}[t!]
    \centering
        \includegraphics[width=18cm]{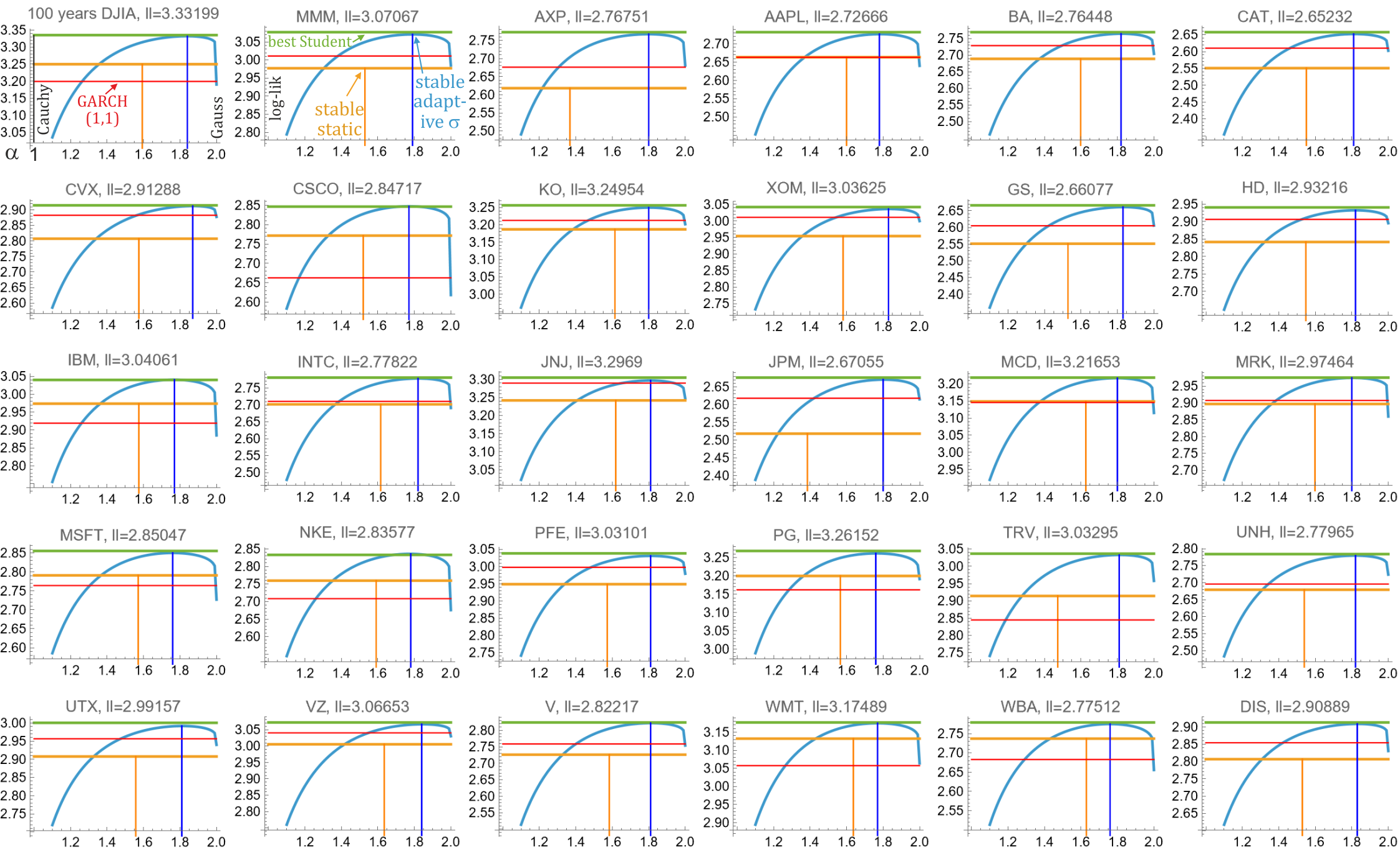}
        \caption{Log-likelihoods (mean $\ln(\rho_t(x_t))$) evaluations for log-returns of 107 years DJIA time series, and 10 years for 29 individual companies. In horizontal axis there is fixed $\alpha$ parameter of stable distribution (from Cauchy to Gauss), assuming parameters (orange), and adaptive $\sigma$ scale parameter (blue, using $p=0.8$ power and $\eta_2=0.05$ learning rate), all for $\mu=0$ center. We can see adaptation has allowed for less heavy tails (larger $\alpha$ in maximum). There are also shown analogously the best from $\sigma$ adaptation for Student's t-distribution~\cite{mystudent} (green). Red line shows evaluation of $\sigma$ adaptation by standard GARCH(1,1) model - which is comparable with $\alpha=2$ Gaussian case.  }
       \label{eval}
\end{figure*}

However, real-life e.g. financial data is usually non-stationary, requiring some \textbf{adaptive estimation} of evolving time-dependent parameters e.g. $\theta_t=(\mu_t,\sigma_t,\alpha_t)$, visualized in Fig. \ref{djiastab}, \ref{finstab}. In classical ARMA-ARCH~\cite{garch} approaches there is assumed some arbitrary dependence between parameters for succeeding times. To avoid their bias, there was proposed \textbf{moving estimator} approach~\cite{adaptive} where we shift estimator to use local past values $\{x_\tau\}_{\tau<t}$ with exponentially weakening weights for the old values:
\be \theta_t=\argmax_\theta F_t\qquad\textrm{for}\qquad F_t = \sum_{\tau<t} \bar{\eta}^{t-\tau} \ln(\rho_{\theta}(x_\tau))\label{opt}\ee
for $\bar{\eta}\equiv 1-\eta$ learning rate, usually above 0.9.  For the final evaluation, also to optimize $\eta$, there is used their average:
\be F = \frac{1}{T}\sum_{t=1}^T f(\theta_t,x_t) \quad \textrm{e.g. log-likelihood: }\frac{1}{T}\sum_{t=1}^T \ln(\rho_{\theta_t}(x_t))\ee

Such separate estimation for each position is computationally costly, e.g. we can minimize (\ref{opt}) with the previous parameters as the starting point for optimization. It can be simplified especially for estimators being averages, which can be modified to moving estimator by just switching to exponential moving average (EMA). For example EPD (exponential power distribution) family with $\rho(x)\sim \exp(-|x|^\kappa)$ type density has  $\sigma^\kappa = E[|x-\mu|^\kappa]$ MLE (maximum likelihood estimator) - changing average to EMA~\cite{adaptive} usually gives better log-likelihood than ARMA-ARCH, also here summarized in Fig. \ref{eval}.

However, usually MLE estimation is much more complex than just averaging, but moments can be easily extended to adaptive with moving averages, and allow to estimate parameters. Working with heavy tails, we need to use moments for low powers as higher become infinite, suggesting to use \textbf{absolute central moments}: $m_p= E[|x-\mu|^p]$ for one or a few powers, which for EMA becomes:
\be m_{p,t+1} = m_{p,t} + \eta (|x_t-\mu_t|^p-m_{p,t}) \ee
and previously was discussed~\cite{mystudent} for moving estimator of Student's t-distribution for all 3 parameters: $\mu,\sigma,\nu$.

Here this approach was adapted for another basic heavy-tailed distribution family: alpha-stable~\cite{stable,stablebook}, obtained e.g. by averaging i.i.d. random variables of not necessarily finite variance in the Generalized Central Limit Theorem~\cite{genCTL}, hence should be often universal for real-life date. 

Stable distribution also leads to $H_q=q/\alpha$ generalized Hurst exponent~\cite{multifractal} for i.i.d time series, describing $\textrm{E}\left(|x_{t+\tau}-x_t|^q\right) \sim \tau^{q H_q}$  behavior, hence the proposed moving $\alpha$ estimator can be also used for adaptive estimation of Hurst exponent.

For dataset of 107 years Dow Jones Industrial Average (DJIA) daily log-returns and 10 years for 29 its recent companies, there was tested such adaptive estimation especially of $\sigma$, leading to essentially better log-likelihood than GARCH(1,1)~\cite{garch}, also than EPD~\cite{adaptive}. 

However, obtained log-likelihood has turned out slightly worse than using adaptive Student's t-distribution~\cite{mystudent}, probably because it covers $\rho(x)\sim |x|^{-\nu-1}$ tails for arbitrarily large power, while stable distribution switches to Gaussian from 3rd power as variance becomes finite.
\section{Time series used for evaluation}
There was used 1900-2007 daily Dow Jones index\footnote{Source of DJIA time series: http://www.idvbook.com/teaching-aid/data-sets/the-dow-jones-industrial-average-data-set/},  working on $x_t=\ln(v_{t+1}/ v_t)$ sequence of daily log-returns.

Figure \ref{eval} additionally contains such evaluation of log-returns for 29 out of 30 companies used for this index in September 2018. Daily prices for the last 10 years were downloaded from NASDAQ webpage (www.nasdaq.com) for all but DowDuPont (DWDP) - there were used daily close values for 2008-08-14 to 2018-08-14 period ($2518$ values) for the remaining 29 companies: 3M (MMM), American Express (AXP), Apple (AAPL), Boeing (BA), Caterpillar (CAT), Chevron (CVX), Cisco Systems (CSCO), Coca-Cola (KO), ExxonMobil (XOM), Goldman Sachs (GS), The Home Depot (HD), IBM (IBM), Intel (INTC), Johnson\&Johnson (JNJ), JPMorgan Chase (JPM), McDonald's (MCD), Merck\&Company (MRK), Microsoft (MSFT), Nike (NKE), Pfizer (PFE), Procter\&Gampble (PG), Travelers (TRV), UnitedHealth Group (UNH), United Technologies (UTX), Verizon (VZ), Visa (V), Walmart (WMT), Walgreens Boots Alliance (WBA) and Walt Disney (DIS).

\begin{figure}[t!]
    \centering
        \includegraphics[width=9cm]{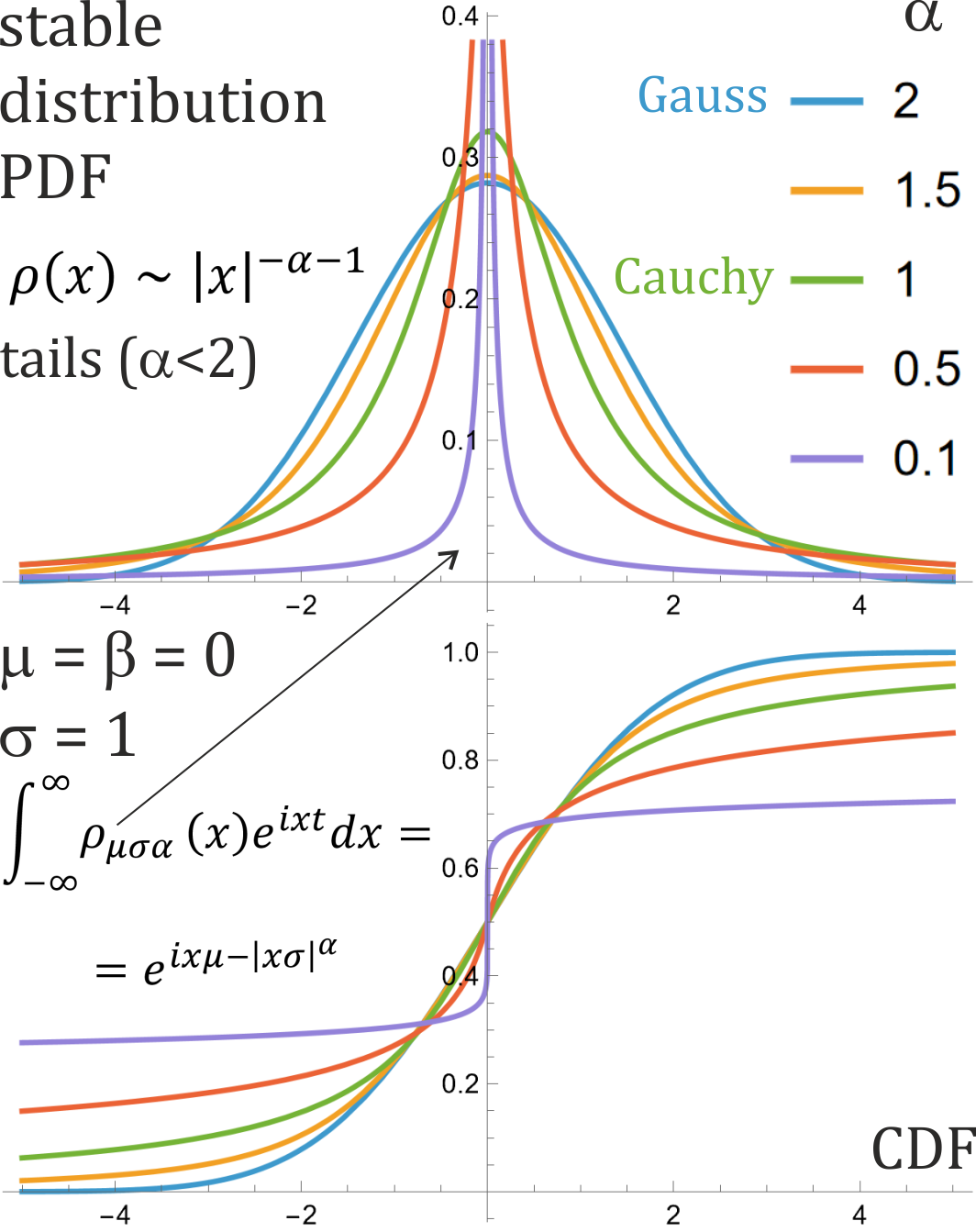}
        \caption{Probability distribution function (PDF, asymptotically $\sim |x|^{-1-\alpha}$) and cumulative distribution function (CDF) for stable distribution with fixed center $\mu=0$ and scale parameter $\sigma=1$, but various shape parameter $\alpha$. We get Gaussian distribution for $\alpha=2$, Cauchy distribution for $\alpha=1$, and can also cover different types of heavy tails and bodies of distribution.}
       \label{stable}
\end{figure}

\begin{figure}[t!]
    \centering
        \includegraphics[width=9cm]{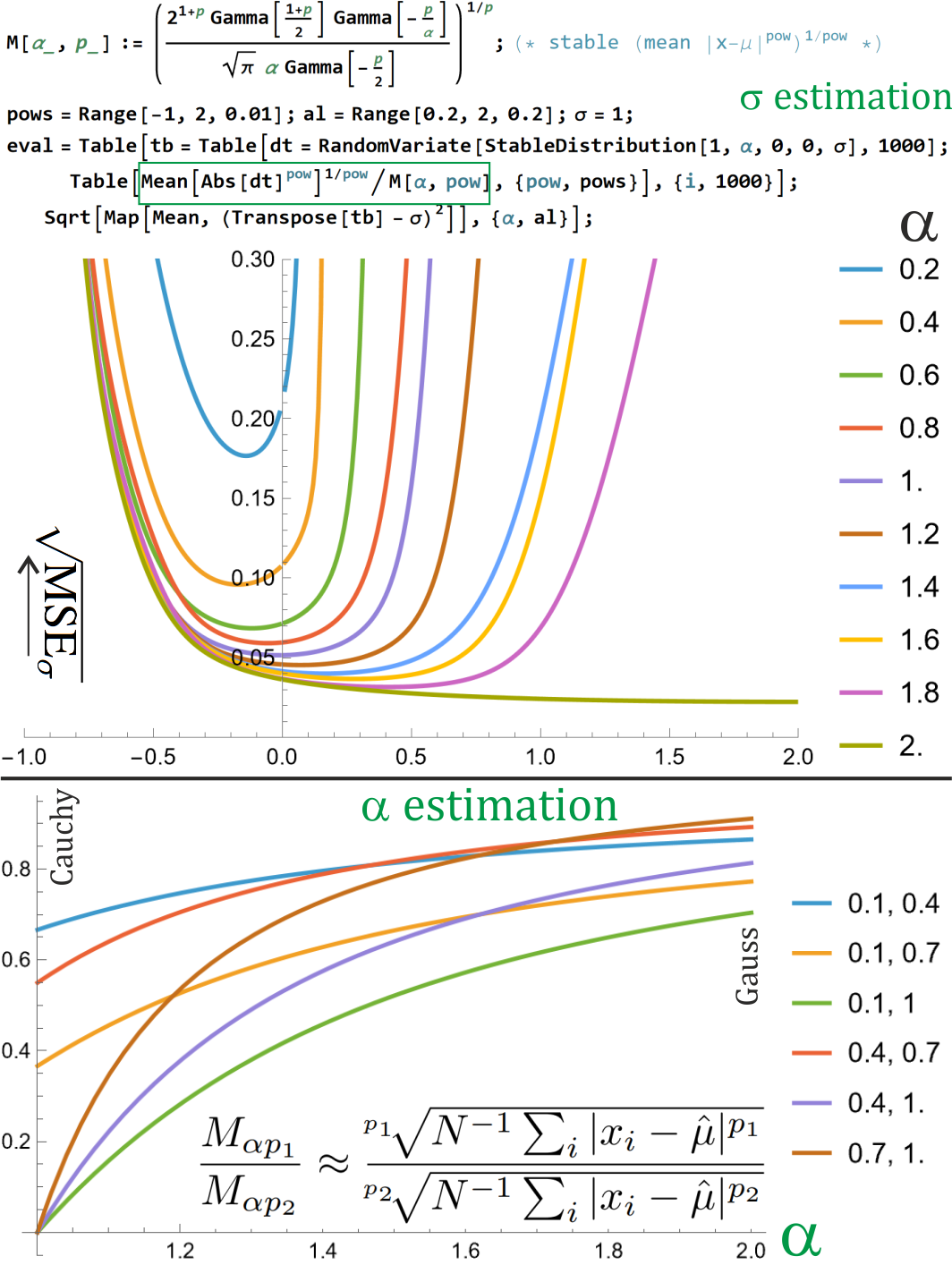}
        \caption{\textbf{Top}: error dependence for choice of power $p$ in $\sigma$ estimation as $\hat{\sigma} =  \sqrt[\leftroot{5}p]{T^{-1} \sum_t |x_i-\hat{\mu}|^p}/M_{\alpha p}$. We can see that for Gaussian distribution $\alpha=2$ we should choose $p=2$ as in standard variance estimation, but to improve prediction should reduce this $p$ for lower $\alpha$, to $\approx (\alpha-1)/2$ for $\alpha\in (1,2)$. \textbf{Bottom}: monotonous functions for $\alpha$ estimation for various choices of 2 powers $p_1,p_2$.  }
       \label{est}
\end{figure}

\section{Stable distribution and adaptation}
A distribution is referred as stable if a linear combination of two independent random variables from this distributions has the same distribution, up location $\mu$ and scale $\sigma$ parameters. They are also referred as L\'{e}vy-stable alpha distributions as probably were  first studied by Paul L\'{e}vy~\cite{stable}.

Beside \emph{location} $\mu$ and \emph{scale} $\sigma$, they also have \emph{stability parameter} $0<\alpha <2$ for $\rho(x)\sim 1/|x|^{\alpha+1}$ probability density of tails (heavy), or $\alpha=2$ being Gaussian distribution with $\rho(x)\sim \exp(-x^2)$ tails.

While generally there can be also discussed further asymmetry/\emph{skewness parameter} $\beta\in [-1,1]$. For simplicity let us start without it: assume $\beta=0$, what leads to a simple characteristic function, with some PDF/CDF shown in Fig. \ref{stable}:

\be \varphi_{\mu \sigma \alpha}(t)=\int_{-\infty}^{\infty} \rho_{\mu \sigma\alpha}(x)\, e^{ixt} dx = e^{ix\mu-|x\sigma|^\alpha} \label{stab} \ee
Characteristic function of sum of independent random variables is product of their characteristic functions: $\varphi_{X+Y}(t)=\varphi_{X}(t) \varphi_Y (t)$, making (\ref{stab}) indeed a stable variable: for the same $\alpha$, location and scale parameter:
\be\mu_{X+Y}=\mu_X +\mu_Y\qquad\qquad \sigma_{X+Y}^\alpha=\sigma_X^\alpha +\sigma_Y^\alpha\ee

It suggests their universality for real-life data, especially through the Generalized Central Limit Theorem~\cite{genCTL}. It says that averaging sequence of independent, identically distributed variables leads to a stable distribution, generalizing the Cental Limit Theorem to infinite variance case.
\subsection{Absolute central moments method}
For method of moments we will use absolute central moments: $E[|x-\mu|^p]$ for not necessarily integer power $p\in \mathbb{R}$. Using Wolfram Mathematica there was calculated moment formula as the below integral (for $\beta=0$), finite for $p<\alpha$:
\be M_{\alpha p}=\sqrt[\leftroot{5}p]{\int_{-\infty}^\infty |x|^p \rho_{01\alpha}(x) dx}=
\sqrt[\leftroot{5}p]
{\frac{2^{1+p}\,\Gamma\left(\frac{1+p}{2}\right)\Gamma\left(\frac{-p}{\alpha}\right)}
{\sqrt{\pi}\,\alpha\, \Gamma(-p/2)}  }\label{mom}
\ee
Having a $\{x_t \}_{t=1..T}$ data sample, fixing $\alpha$ and using some $\mu$ estimator e.g. approximate $\hat{\mu}=T^{-1} \sum_t x_t$ as just mean, the above formula gives simple estimator of scale parameter $\sigma$:
\be \hat{\sigma} =  \frac{\sqrt[\leftroot{5}p]{T^{-1} \sum_t |x_t-\hat{\mu}|^p}}{M_{\alpha p}} \label{sigmaest}\ee
The used $p$ has to be in $(-1,\alpha)$ range, where the possibility to use non-integer $p$ might be crucial for the $p<\alpha$ requirement.

Additionally, using various $p$ for such $\sigma$ estimation has different uncertainty depending on $\alpha$, as shown in Fig. \ref{est} - suggesting to optimize $p$ e.g. based on the used $\alpha$ range, or even modify $p$ dynamically. For $\alpha=2$ the optimal $p$ is $2$ variance estimation. For $\alpha\in (1,2)$ the optimal $p$ is $\approx (\alpha-1)/2$.\\

To estimate $\alpha$, a natural direct way is to divide such averages for two different powers $p_1,p_2$, removing $\sigma$ dependence:
\be \frac{M_{\alpha p_1}}{M_{\alpha p_2}} \approx
\frac{\sqrt[\leftroot{7}p_1]{N^{-1} \sum_i |x_i-\hat{\mu}|^{p_1}}}
{\sqrt[\leftroot{7}p_2]{N^{-1} \sum_i |x_i-\hat{\mu}|^{p_2}}}\label{nuest}\ee
Choosing some $p_1\neq p_2$, the $M_{\alpha p_1}/M_{\alpha p_2}$ is monotonous with $\alpha$ (examples in Fig. \ref{est}), we can e.g. put its behavior into a table and interpolate based on the averages to estimate $\alpha$, e.g. done as \verb"find"$\alpha$ in the code in Fig. \ref{djiastab}.

The finally found parameters might be worth tuning by some small modification, e.g. in Fig. \ref{djiastab} there was used $\sigma m=1.05$ to multiply the final $\sigma$, allowing to slightly improve log-likelihood.

\subsection{Moving central moments estimators}
Above methods of moments can be easily adapted for moving estimator by just replacing averages with exponential moving averages - uniform weights with exponentially weakening.

For the center $\mu$ we can use just a basic adaptation below - it is optimal only for the Gaussian case ($\nu\to \infty$), hence generally it could be slightly improved. However, for the discussed data the gains were already nearly negligible.
\be \mu_{t+1} = \mu_{t} + \eta_1 (x_t -\mu_t ) \ee

\begin{figure}[t!]
    \centering
        \includegraphics[width=9cm]{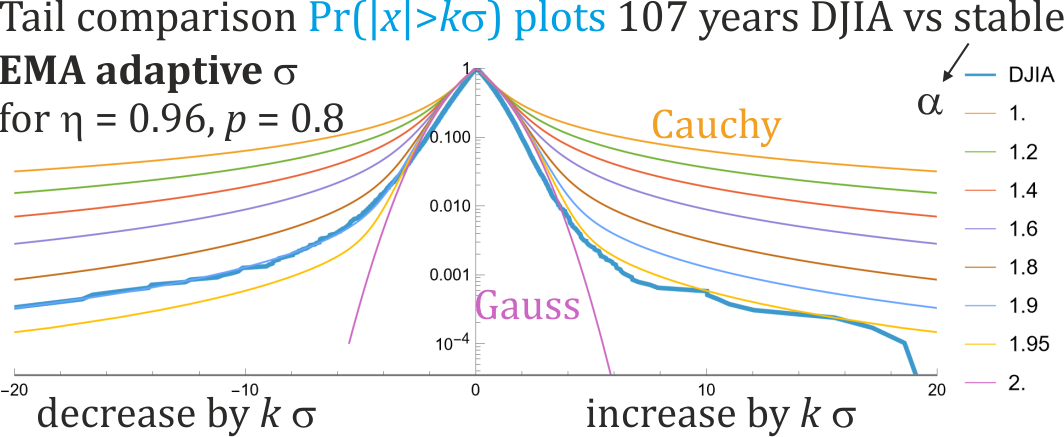}
        \caption{Tail visualization: probability of exceeding $k\sigma$ toward left (negative) and right (positive): based on DJIA data (bold blue), and its comparison with of stable distribution for various $\alpha$ parameter (thin color lines). We can observe negative tail fits well $\alpha=1.9$, positive tail $\alpha=1.95$.}
       \label{tail}
\end{figure}

The most crucial is $\sigma$ scale parameter adaptive estimation, as e.g. in ARCH family but in more agnostic way, here using (\ref{sigmaest}) formula for a chosen $p\in (-1,2)$ power ($p<\min_t(\alpha_t)$), this time with (central absolute) moments evolving in time:
\be m_{p,t+1} = m_{p,t} + \eta_2 (|x_t-\mu_t|^p-m_{p,t}) \ee

Finally, for $\alpha$ estimation we can use (\ref{est}) formula for analogously updated moments for some 2 different powers $p_1, p_2$ and some $\eta_3$ learning rate. 

Figure \ref{djiastab} contains used Mathematica code for adaptation of all 3 parameters, with their evolution for DJIA time series. Manual tuning has lead to 3 different learning rates there: $\eta_1=0.002, \eta_2=0.03, \eta_3=0.006$ for correspondingly $\mu,\sigma,\alpha$ (much faster evolution of volatility/scale parameter $\sigma$).

Figure \ref{eval} shows evaluation using fixed $\mu=0$ center and various fixed $\alpha$ for single MLE $\sigma$ parameter, or $\sigma$ adapted using (\ref{sigmaest}) estimation with $p=1$ power and $\eta_2=0.05$ learning rate, just slightly worse that for Student's t-distribution from \cite{mystudent}. 

The $\alpha$ evolution, unavailable in standard ARMA-ARCH approaches, evaluates local tail shapes, probability of potentially destabilizing extreme events - allowing to indeed interpret it as \textbf{stability}, complementing popular \textbf{volatility} evaluation similar to $\sigma$. For example it shows that in 1967-1983 there was nearly Gaussian distribution, also confirmed in \cite{mystudent}. We can  observe events like Black Monday leading to sudden drops of market stability. Having evolving evaluation of market stability brings a chance to understand various factors influencing it, and maybe try to control them.
\subsection{Including asymmetry}
The standard approach to include asymmetry for stable distribution is through $\beta\in[-1,1]$ \emph{skewness parameter} - modifying the characteristic function for $\alpha \in (1,2]$ to:
\be\varphi_{\mu \sigma \alpha}(t)= \exp\left(ix\mu-|x\sigma|^\alpha (1-i\beta \textrm{sgn}(x)\tan(\pi \alpha/2))\right) \ee
It might be worth to include its estimation, what could be done e.g. by extending to EMA updated moments maintaining sign: $E[\textrm{sgn}(x)|x|^p]$. However, just fixing $\beta$ can already bring slight improvement of log-likelihood, like $\beta=-0.3$ in Fig. \ref{djiastab}.\\

Alternative general approach, e.g. mentioned for Student's t-distribution \cite{mystudent}, and suggested by tail analysis in Fig. \ref{tail}, is just gluing two distributions of slightly different parameters - to be used for  $(x_t-\mu_t)/\sigma_t$ online normalized variables.

For example, as $\alpha$ in Fig. \ref{tail} turns out $\Delta\approx 0.05$ lower for $x<\mu$ negative tail, we can estimate $\alpha$ as for $x>\mu$ positive tail, and use $\alpha-\Delta$ for the negative tail. Working on $(x_t-\mu_t)/\sigma_t$ normalized values, we need probability density for normalized $\mu=0,\sigma=1$, or approximated and potentially different for left/right tails: $\sigma_l,\sigma_r \approx 1$. Finally, including continuity for $x=\mu$, we get such asymmetric PDF as:
\be
\frac{2}{\frac{\sigma_l}{\rho_0(\alpha)}+\frac{\sigma_r}{\rho_0(\alpha-\Delta)}}
\left\{
\begin{array}{cc}
 \frac{\rho_{0,1,\alpha}(x/\sigma_l)}{\rho_0(\alpha)}\quad  \text{ if }x\leq \mu  \\
 \frac{\rho_{0,1,\alpha-\Delta}(x/\sigma_r)}{\rho_0(\alpha-\Delta)}\quad \text{ if }x> \mu \\
\end{array}
\right.
\ee
for $\rho_0(\alpha)=\rho_{0,1,\alpha}(0)=\Gamma(1+1/\alpha)/\pi$ density in $\mu$. Unfortunately for DJIA data it did not allowed to improve log-likelihood.
\section{Adaptive Hurst exponent, heavy tail removal}
There is popular estimation of multifractal Hurst exponent, for example as $\textrm{E}\left(|x_{t+\tau}-x_t|^q\right) \sim \tau^{q H_q}$ behavior for $H_q$ generalized Hurst exponent.

While its main  motivation is evaluation of long-range behavior, for heavy tails already for i.i.d. variables it obtains e.g. $H_q=q/\alpha$ for $\alpha$ parameter stable distribution - allowing to estimate it in adaptive way through moving estimator of $\alpha$.

Moreover, the proposed adaptive estimation allows to include in considerations, or even remove the heavy tail dependence - allowing for \emph{compensated Hurst exponent} indeed focused only on evaluation of long range behavior. For this purpose, we can for example normalize the variables by transforming with estimated CDF of stable or Student's t-distribution, and then inverted CDF of target distribution e.g. $N(0,1)$ normalized Gaussian:
 $$ x_t \to \textrm{CDF}^{-1}_{N(0,1)}(\textrm{CDF}_{\theta_t}(x_t))$$ then calculating Hurst exponent for such normalized variables.

\section{Conclusions and further work}
This article extends the method of moments to adaptive case for EMA moving estimation of $(\mu_t,\sigma_t,\alpha_t)$ parameters of alpha-stable distribution analogously to previous articles for exponential power distribution~\cite{adaptive} and Student's t-distribution~\cite{mystudent}.

\begin{figure}[t!]
    \centering
        \includegraphics[width=9.cm]{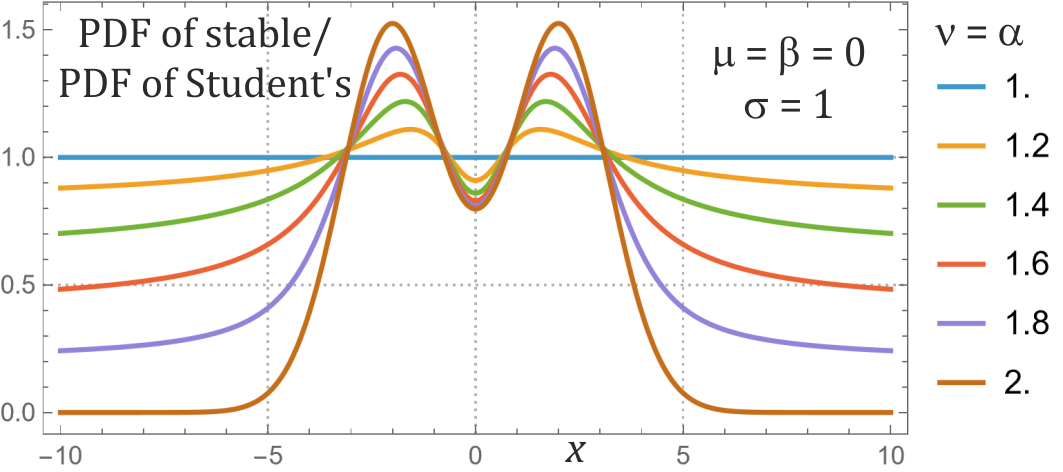}
        \caption{Division of PDF of stable by Student's t-distributions for $\mu=\beta=0$, $\sigma=1$ and varying $\alpha=\nu$ deciding power of 1/polynomial tail.}
       \label{div}
\end{figure}

Figure \ref{tail} suggests stable distribution is appropriate to model tails of financial data, however, the obtained log-likelihood was slightly worse than for Student's t-distribution, probably due to covering 1/polynomial tails only up to degree 3 or their differences visualized in Fig. \ref{div}, suggesting to consider some hybrid between Student's t and stable distribution, what is planned for further work, e.g. just switching to Student for when evolving $\alpha$ exceeds 2, or maybe automatically optimizing shape of parametrized distribution based on the data.

Adaptive estimation of $\alpha$ distribution allows to continuously evaluate market stability, e.g. to understand and try to control it, react to crucial events. Also can be used as additional variable to improve predictions of various models.

Such found evolving distribution can be used as initial stage e.g. by normalization to nearly uniform distribution in $[0,1]$ by $x_t\to \textrm{CDF}_{\theta_t}(x_t)$ for further analysis e.g. of joint distributions, or long-range dependencies e.g. with Hurst exponent.

Examples of plans for further work:
\begin{itemize}
  \item Extension of stable distribution to $\alpha\geq 2$ with $\rho(x)\sim |x|^{-\alpha-1}$ in a continuous way to include adaptation for one over polynomial tails of power 3 and higher, e.g. by switching to Student's t-distribution, or some their combination, or automatically optimized based on the data.
  \item Adaptation of $\beta$ parameter and development of alternative approaches to asymmetry.      
  \item Improve estimators from moments - especially of $\alpha$.
   \item Add further modelling, like dependence from other stocks, macronomical data, e.g. with adaptive linear regression~\cite{regression}, and HCR~\cite{hcr} to include subtle dependencies.
  \item Find alternative approaches for moving estimators of various distributions, e.g. with gradient ascend approaches, maybe also including 2nd order information like in \cite{OGR}.
  \item The discussed approach has many hyperparameters like learning rates -  often universal for similar data types. It might be valuable to automatically optimize them, adapt through evolution.
  \item Understand mechanisms/dependencies affecting $\alpha$ evolution, also separate for left/right tail, and hopefully exploit them e.g. to improve market stability.
  \item Test discussed approaches for different application like data compression, where log-likelihood improvement translates into nit/symbol savings.
\end{itemize}

\bibliographystyle{IEEEtran}
\bibliography{cites1}
\end{document}